# Heterogeneity and anomalous critical indices in the aftershocks distribution of L'Aquila earthquake

**D. Innocenti (^), M. Iovenitti (^), N. Poccia (^), A. Ricci(^), M. Caputo (^)(\*\*), A. Bianconi (^)\***

(^) Department of Physics, Sapienza University of Rome, P. le A. Moro 2, 00185 Roma, Italy,

(\*\*) Department of Geology and Geophysics, Texas A&M University, College Station, TX, 77843, USA.

\*E-mail: antonio.bianconi@roma1.infn.it

**Abstract.** The data analysis of aftershock events of L'Aquila earthquake in Apennines following the main 6.3 Mw event of April 6, 2009 has been carried out by standard statistical geophysical tools. The results show the heterogeneity of seismic activity in five different geographical sub-regions indicated by anomalous critical indices of power law distributions: the exponents of the Omori law, the b values of Gutenberg-Richter magnitude-frequency distribution, and the distribution of waiting times. The heterogeneous distribution of dynamic stress and a different morphology in the 5 sub-regions has been found and two anomalous sub-regions have been identified.

## 1. Introduction.

In these last ten years advances, first, in fast electronic data collection in different fields like geophysics, astrophysics, material science, biology and, second, in the distribution via world wide web of a large amount of data in open large data-banks have allowed the physics community to develop the field of data analysis of complex natural phenomena opening new perspectives for theoretical physics looking at simple laws. The analysis of the large data set on earthquakes shows power law distributions [1-3] in common with many other complex phenomena [4,5] that are attracting a large interest since this behavior in nature reflects the tendency of large systems with many components to evolve into a critical state showing simple physical laws. While the power law behaviors look universal the actual exponents of the power law distributions reflect the distribution of dynamical and structural features in the system. Moreover many systems are made of multiple modules so that the scale invariance is not respected. Therefore the scientific community is focusing on identifying the heterogeneity in complex phenomena beyond the simple average universal behavior.

The earthquakes show universal simple power law distributions as shown by the intensities of earthquakes occurring in California between 1910 and 1992, [1] as well as in Italy between 1000 and today [2,3], the Gutenberg-Richter's magnitude-frequency distribution of aftershocks, and the spatial distribution of aftershocks the exponents of the Omori law [6], the distribution of waiting times [7] and the radial distribution function [8]. However beyond the universal behavior the earthquakes show relevant heterogeneity, as it has been suggested from observations of the relation between natural fault and earthquake models [9]. Recently, multiscale heterogeneous distribution of the dynamic stress energy has been proposed to explain the diversity of earthquake sequences [10,11]. Earthquake faults occur in topologically complex, multiscale networks or systems that are driven to failure by external forces arising from plate tectonic motions [12-14].

The multiscale heterogeneous picture is due to the spatial and temporal scales of the problem. The spatial scales range from the microscopic scale (1μm to 1 cm) associated with friction to the tectonic plate boundary scale ($10^3$–$10^4$ km) associated with the driving force. The temporal scales range from seconds (during dynamic stress) to $10^3$–$10^4$ years (repeat times for earthquakes) to $10^7$–$10^8$ years (evolution of plate boundaries). Research is now focusing on a statistical physics approach [15] with the aim to understand the implications of space-time correlations and dynamical patterns in these fundamentally multiscale phenomena related with the fractal topology of fault systems and show up in the linearity of the Gutenberg-Richter magnitude-frequency relation, and in other observed statistics of earthquakes, including the Omori aftershock relation.

The recent L'Aquila earthquake in Italy has inspired this work. Italy is a well-known heterogeneous earthquake zone, it's not a simple area like AS the West Coast of California where there are two large plates sliding against each other along the San Andreas fault. Because of the African and European plates collision, the Apennine chain is highly fractured and broken up, with a lot of microplates moving around, which creates a lot of different types of fault action. In order to more understand the dynamical mechanism which triggers aftershocks in Italy Apennine, we have investigated the heterogeneity of the critical exponents of Gutenberg-Richter magnitude-frequency relation, the Omori aftershock relation and the distribution of waiting times of the seismic events following the 6.3 Mw L'Aquila earthquake of April 6, 2009. We have identified five different sub-regions (inside each regions the exponents are uniform) and we report that in one of these regions they show largely anomalous values.

2. **Results and discussion.**

We have taken the earthquakes data covering the period 01/01/2008 – 04/24/2009 in a region spanning 42.0°N-42.6°N latitude and 12.0°E-13.0°E longitude from 'Italian Seismic Bulletin' of the Italian Seismic Instrumental and parametric Data-base (ISIDE) available on the web site [16,17]. The distribution of epicenters is plotted in Fig. 1. The transformation of polar coordinates (latitude, longitude) in Cartesian coordinates (kilometers) has been done using the average radius of the Earth (r = 6371 km). We have used the interferogram measured by a satellite to locate events in the Cartesian space on a geographical map of the region [18,19]. The ENVISAT interferogram over the L'Aquila area is obtained by INGV scientists using the technique known as SAR Differential Interferometry (DInSAR) on satellite radar data from ESA's ENVISAT and the Italian Space Agency's COSMO-SkyMed. ENVISAT interferogram has been calculated from the difference between a pair of images taken on February 1, 2009 and April 12, 2009. The ENVISAT interferogram shows nine concentric fringes that identify the area of maximum displacement where the ground has shifted about 25 cm along the direction of view of the satellite. We have superimposed the aftershock events on the interferogram in order to visualize the events spatial distribution. The first thing one observes is that the main event does not fall in the Center of the nine Concentric Fringes of the Interferogram (CCFI).

The origin in our map is placed at the CCFI which is located at 42.323° N and 13.462° E, while the main shock (Mw 6.3) that occurred at 01:32:39 (UTC) on April 06, 2009 has been located by INGV at 42.334° N and 13.334°E, and depth 8.8 km.

We have addressed our data analysis search to identify different sub-regions that show uniform power law distributions for Gutenberg Richter and Omori Laws. We have identified five groups of aftershock events that occur in five sub-regions identified with different colors in Fig. 1. The nine concentric fringes of the interferogram contain only three (1+2, 3 and 4) of the five sub-regions as it is shown in Fig. 1. We observed that the main event falls on the edge of the fringes and is located in an area with a low density of aftershocks. We have calculated the centroid of the distribution of aftershock events and we have found that it occurs very close the CCFI.

We have calculated the radial distribution function taking as center the CCFI in two different time intervals (the first from April 6 to April 12 and the second from April 6 to April 24). The distribution remains almost unchanged and it is reported in Fig. 2. If the location of the 6.3 Mw main event is taken as the center of the distribution function it shows very large changes in time.

Looking at the data it is clear that the spatial distribution of events is not homogeneous and seven sub-regions can be easily recognized. The seven sub-regions have been studied by the methods of statistical physics (Omori Law, Gutenberg-Richter's magnitude-frequency distribution, waiting time distribution) and the results obtained show that the all the seven sub-regions can be grouped into five sub-regions (1+2, 3, 4, 5, 6a+6b) delimited by the points reported in Table 1.

The first evidence that the system is heterogeneous is indicated by the fact that the five sub-regions show different values of the b coefficient of the Gutenberg-Richter law. A similar spatial heterogeneity is indicated by b variation and it has been found recently near the mount Etna in Sicily [20]. In Fig. 3 we report the cumulative distribution of the magnitude of the after-shocks in the five sub-regions or zones that follow the Gutenberg-Richter law. For each zone the value of b has been calculated using different cut-off magnitude ($M_c$). We have considered all the events above magnitude $m_0 = 1.4$ up to a variable maximum magnitude in the range $3<M_c<4.5$, as it is shown in box inside Fig. 3, with the aim to estimate the errors on the b values. The high heterogeneity between the five zones is clearly indicated by the high variation of b values. The b values of two sub-regions (1+2) and 4 are larger than those of any other zones.

The investigation of the Omori Law in the five sub-regions can provide information on the different dynamics. For each area we have identified a main event that is chosen as the time origin of the aftershocks. The exponents α of the Omori law for zones 3, 5, (6a+6b) have been found to be 0.61, 0.50, 0.89 respectively, as shown in Fig. 4. The large variation in the α values indicate different morphology and dynamic stress relaxation in the different zones.

The regions (1+2) and (4) show a large deviation from the expected Omori law. The cumulative distribution of the Omori law, reported in Fig. 5, highlight the deviation from the logarithmic expected behavior (α=1) [21].

Fig. 5 shows that in the sub-regions (1+2) and 4 the time evolution of aftershocks is clearly anomalous, as indicated in particular by the linear behavior in the (1+2) sub-region.

The earthquakes are related in space and time so we measure a distribution function of the intercurrence time T between earthquakes, $P_{I,L}$ (T) that obeys a unified scaling law expressing a specific organization in time, space and magnitude [7]. The distribution of waiting times T, between earthquakes with magnitude $m = \log_{10}(I)$, take account of a spatial subdivision of the area with a grid of linear size.

In Fig. 6 we report the distribution function $P_{I,L}$ (T) as a function of intercurrence or waiting time [5,20] T, for the (1+2) and 3 sub-regions. We notice that the sub-region 3 exhibits a linear region, where the power law distribution indicates a correlated regime, and a non-linear one that provides evidence of an uncorrelated regime. Indeed for the anomalous sub-region (1+2) we observe only the linear regime, without the crossover to the uncorrelated one.

## 3. Conclusion.

We have analyzed the after shock data showing power law distribution of the 6.3 Mw L'Aquila earthquake. Earthquakes are an important part of the relaxation mechanism of the crust which is submitted to inhomogeneous increasing stresses accumulating at continental-plate borders. The Gutenberg-Richter magnitude-frequency relation, the Omori aftershock law and the waiting time distributions, that probe the state of marginal stability and near-criticality of the system have been investigated. [21-25]. The radial distribution function of aftershocks is centered at the CCFI and not on the main event.

We have identified five different zones with different exponents of the power law distributions. That depends on the morphology of the zone. Finally our analysis has shown clearly the presence of two sub-regions (1+2) and 4 which are quite anomalous, that could be caused by the multiscale heterogeneous distribution of the dynamic stress accumulated after the main event.

**Acknowledgments**

Thanks are due to prof. Giorgio Fiocco for helpful discussions and dr. Fabio Ferrarotto for help. The research has been supported by Sapienza University of Rome.

**Table 1**

|           | $(x_1, y_1)$ [km] | $(x_2, y_2)$ [km] | $(x_3, y_3)$ [km] | $(x_4, y_4)$ [km] | $(x_5, y_5)$ [km] |
|-----------|-------------------|-------------------|-------------------|-------------------|-------------------|
| **Zone1+2**   | (3.79, 5.44)      | (-6.65, 2.89)     | (-1.27, -12.54)   | (8.86, -5.53)     |                   |
| **Zone 3**    | (-0.04, 2.59)     | (-17.85, 14.65)   | (-17.54, 1.49)    | (-6.65, -2.89)    |                   |
| **Zone 4**    | (-0.04, 2.59)     | (5.88, 9.63)      | (-1.58, 9.17)     | (-16.47, 17.72)   | (-19.23, 15.75)   |
| **Zone 5**    | (-16.47, 17.72)   | (-1.58, 9.17)     | (5.88, 9.63)      | (13.11, -0.96)    | (-14.80, 18.60)   |
| **Zone 6a+6b**| (3.47, 28.90)     | (-20.00, 26.27)   | (5.88, 9.63)      | (13.11, -0.96)    | (-14.80, 18.60)   |

**Table 1.** The coordinates (x, y) in Fig. 1 of the vertexes of the polygons that define the five identified sub-regions.

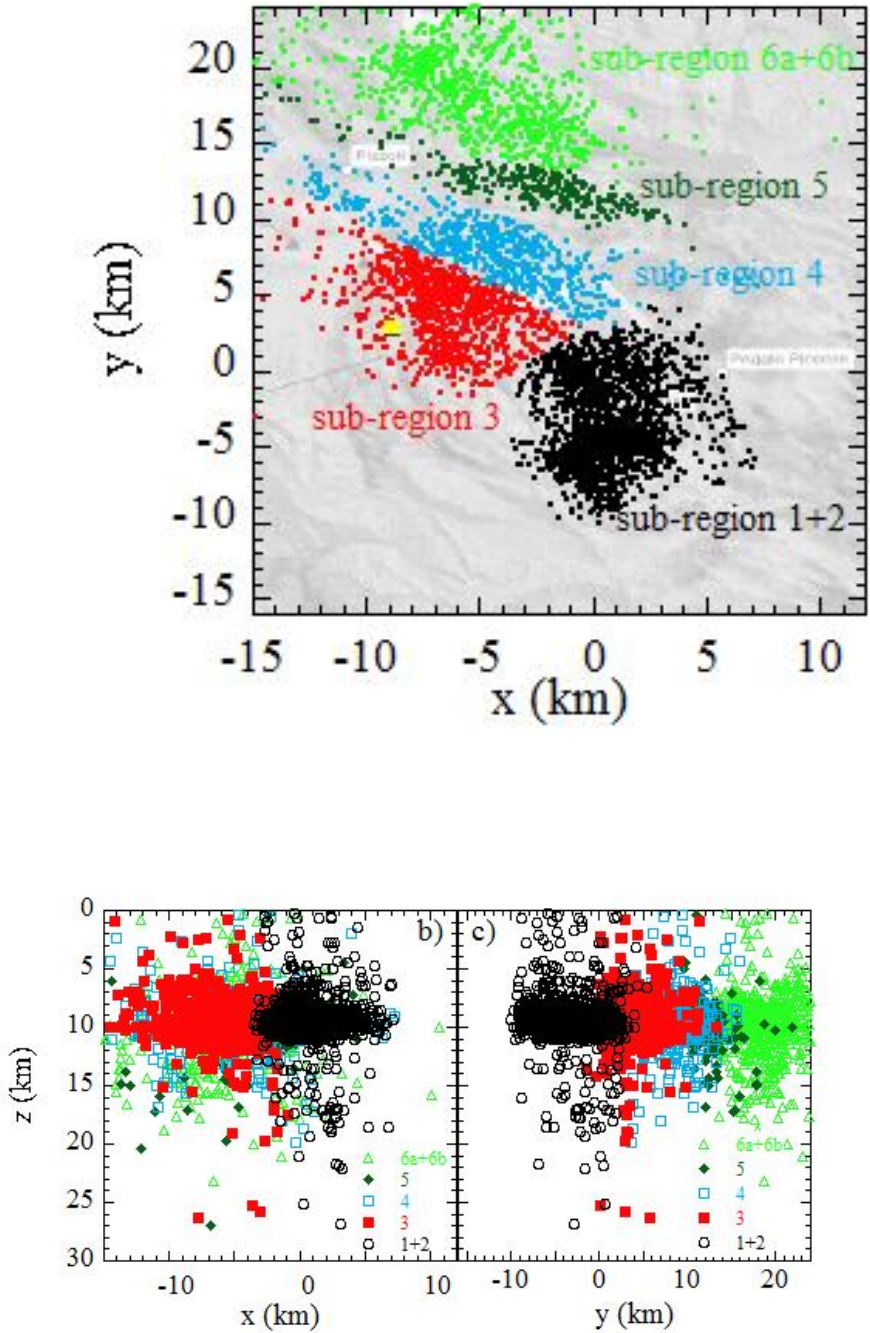

**Fig. 1.** Cross-sections projection on planes (x,y), (x,z) and (y,z) of the aftershocks of 6.3 Mw main shock. (a) Epicenter distribution of aftershocks as shown on the ENVISAT interferogram. (b) The cross-section projected on the (x,z) plane. (c) The cross-section projected on the (y,z) plane. Data here are for 18 days of aftershocks of main shock. The five different sub-regions we have acknowledged

are sub-region (1+2) (open black circles), sub-region 3 (filled red square), sub-region 4 (open blue squares), sub-region 5 (filled dark green diamonds) and sub-region 6a+6b (filled light green circles).

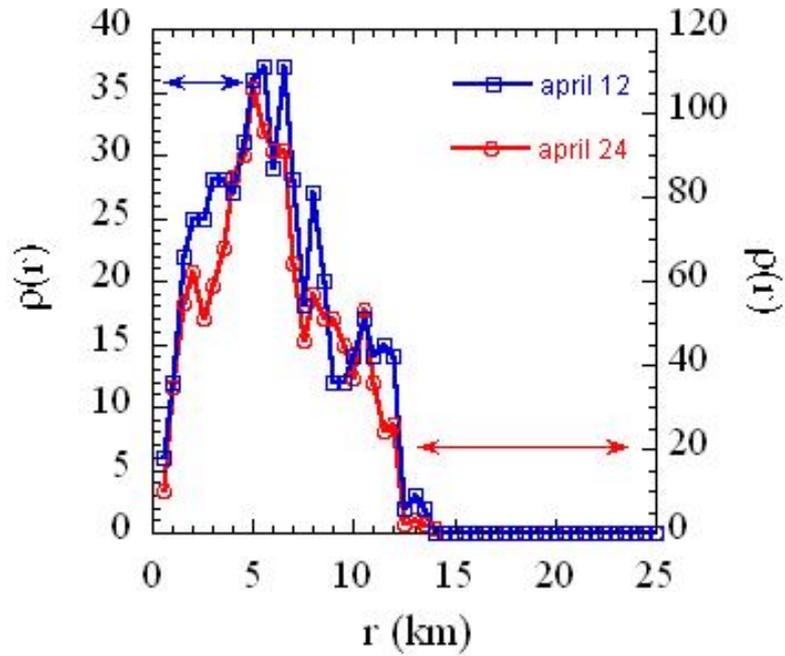

**Fig. 2.** The radial distribution functions of aftershock events taking as the center the CCFI calculated for two different time periods: from 04/06/2009 to 04/12/2009 (april 12 curve) and from 04/06/2009 to 04/24/2009 (april 24 curve).

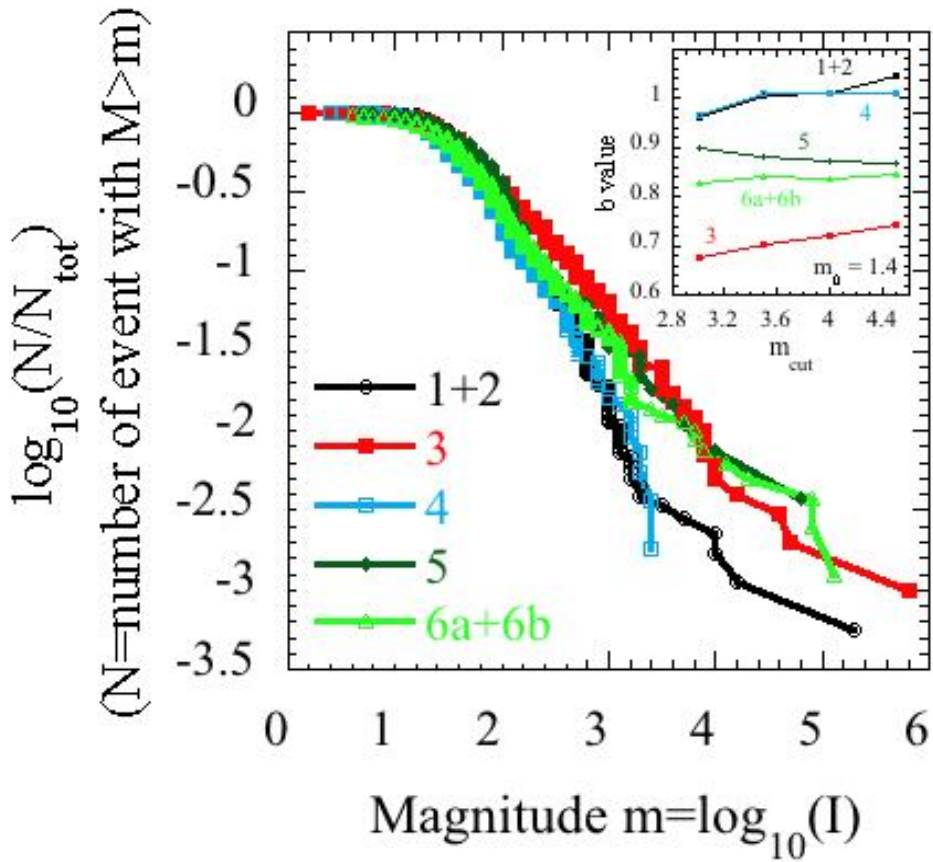

**Fig. 3.** Plot of the logarithm number of aftershocks $N$ versus magnitude $m$ for each zone. $N_{tot}$ is the total number of aftershocks in each sub-regions. An aftershock is an earthquakes that occurs after previous earthquake (the main shock). The insert box reports the $b$ value fit as a function of magnitude cut. For each zone, we fit the logarithm number of aftershocks $N$ versus magnitude $m$ using the Gutenberg-Richter law $\log_{10} N = A - bm$, where $A$ is a constant; we chose different cuts for magnitude greater than $m = 1.4$ (from $m = 3.5$ to $m = 4.5$ increasing $m$ by step of $0.5$) and plot the fit.

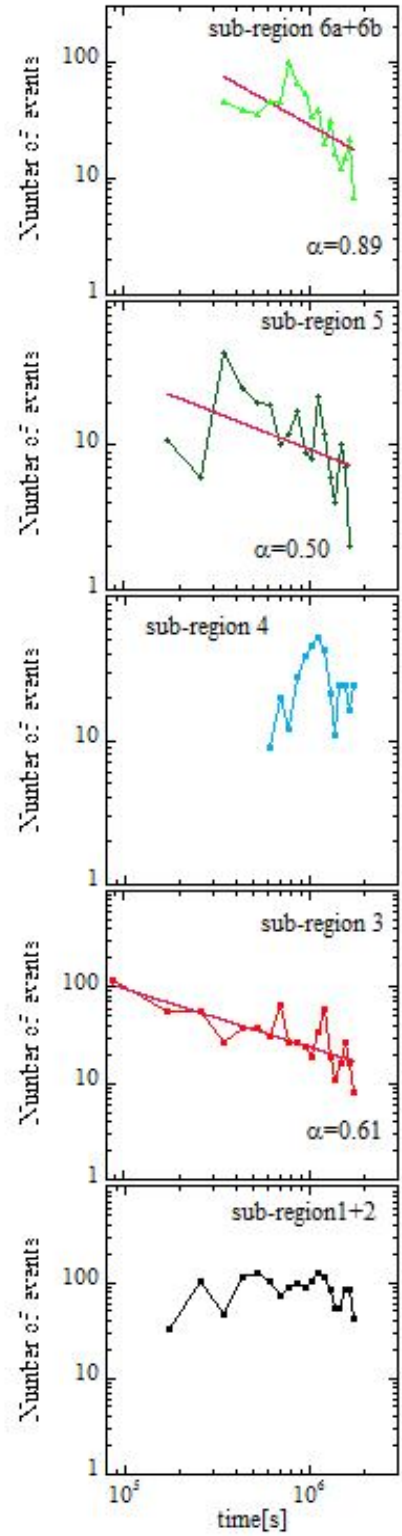

**Fig. 4.** In this figure we represented the daily number of aftershocks *n (t)* in terms of time for each zone. The origin of time is fixed at the proper main event of each zone. For the 3, 5 and (6a+6b) sub-regions, we have studied the data with the Omori law $n(t) \propto T^{-\alpha}$, where $\alpha$ modifies the decay rate and typically falls in the range 0.7-1.5. According to this equation, the rate of aftershocks decreases quickly with time.

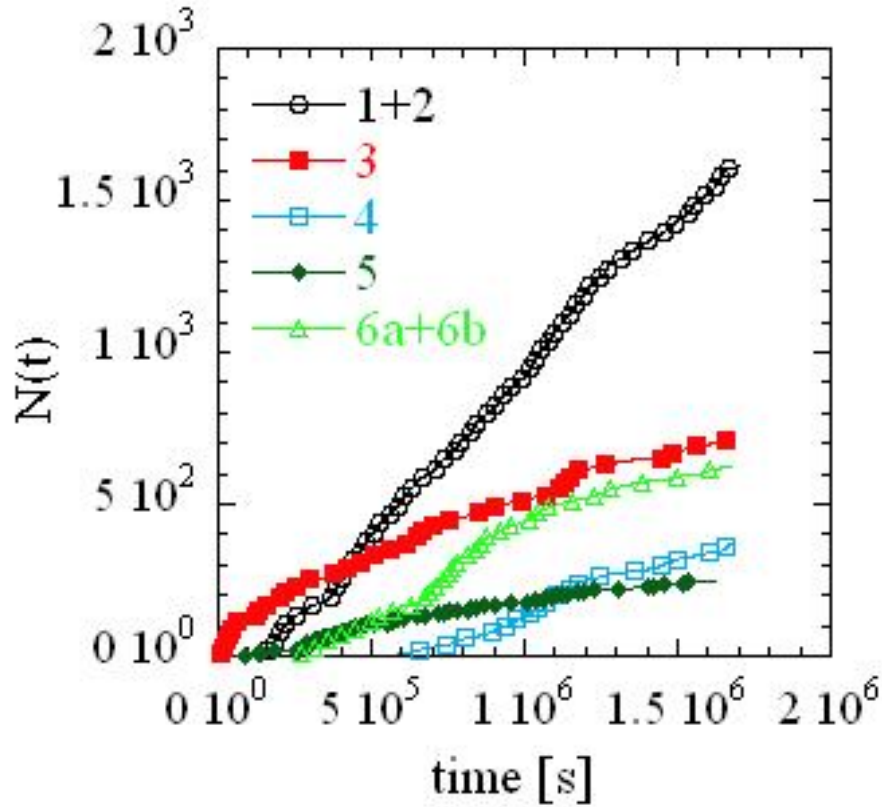

**Fig. 5.** The cumulative distribution of the number of the aftershocks *N (t)* occurred until time *t* after the proper main shock of each zone, defined as integral of *n (t), a*ccording to Omori the expected behavior for $\alpha = 1$ is a logarithm behavior.

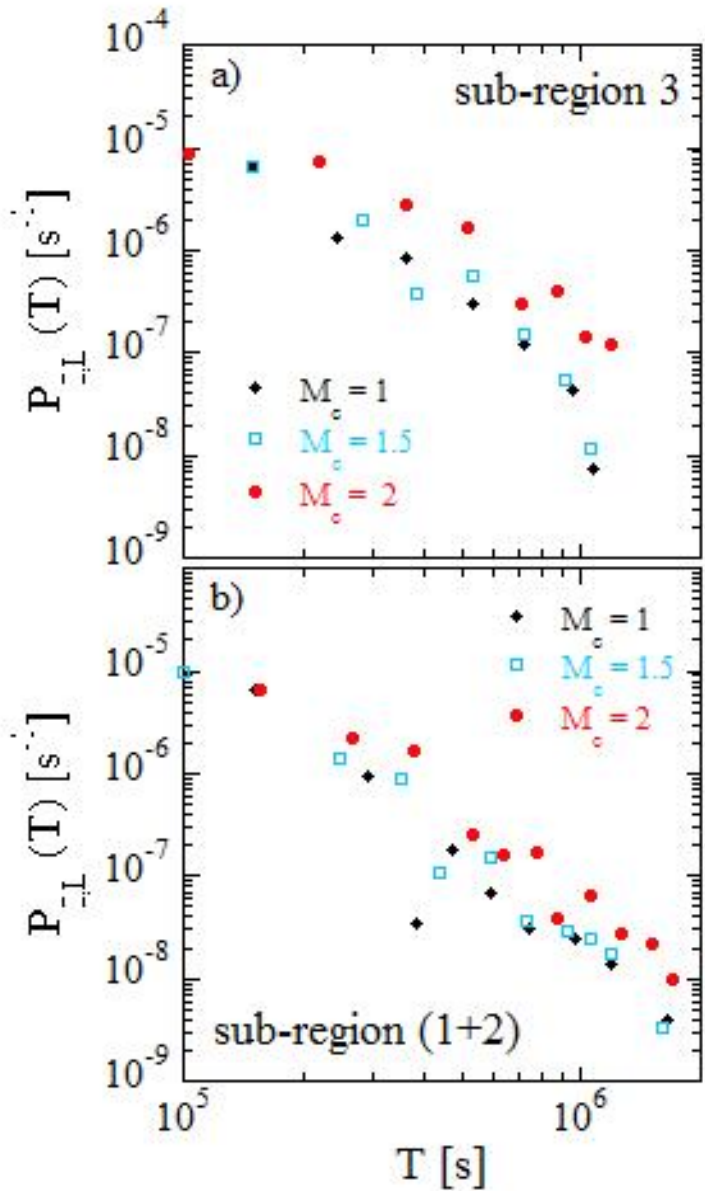

**Fig. 6.** The distribution $P_{I,L}(T)$ as a function of intercurrence time T for the sub-zone 3 (a) and the sub-zone (1+2) (b). The distributions are plotted with fixed linear size L = 1 km for three different cut-off m = 1 (black filled diamonds) 1.5 (blue open squares) and 2 (orange filled circles).